\documentclass[%
 aip,
 rsi,
 amsmath,amssymb,
 reprint,%
]{revtex4-2}

\usepackage[colorlinks=true, allcolors=blue]{hyperref}
\usepackage{graphicx}
\usepackage{dcolumn}
\usepackage[alsoload=synchem]{siunitx} 
\newcommand{\rom}[1]{\uppercase\expandafter{\romannumeral #1\relax}}

\usepackage[utf8]{inputenc}
\usepackage[T1]{fontenc}
\usepackage{mathptmx}

\begin{document}

\preprint{AIP/123-QED}

\title{Beam power scale-up in MEMS based multi-beam ion accelerators}

\author{Q. Ji}
\email{QJi@lbl.gov}
\affiliation{Acceleration Technology \& Applied Physics, Lawrence Berkeley National Laboratory, 1 Cyclotron Road, CA 94720, USA}

\author{K. K. Afridi}
\affiliation{Cornell University, Ithaca, New York, USA}

\author{T. Bauer} 
\affiliation{Acceleration Technology \& Applied Physics, Lawrence Berkeley National Laboratory, 1 Cyclotron Road, CA 94720, USA}

\author{G. Giesbrecht}
\affiliation{Acceleration Technology \& Applied Physics, Lawrence Berkeley National Laboratory, 1 Cyclotron Road, CA 94720, USA}

\author{Y. Hou} 
\affiliation{Cornell University, Ithaca, New York, USA}

\author{A. Lal}
\affiliation{Cornell University, Ithaca, New York, USA}

\author{D. Ni}
\affiliation{Cornell University, Ithaca, New York, USA}

\author{A. Persaud}
\affiliation{Acceleration Technology \& Applied Physics, Lawrence Berkeley National Laboratory, 1 Cyclotron Road, CA 94720, USA}

\author{Z. Qin}
\affiliation{Acceleration Technology \& Applied Physics, Lawrence Berkeley National Laboratory, 1 Cyclotron Road, CA 94720, USA}

\author{P. Seidl}
\affiliation{Acceleration Technology \& Applied Physics, Lawrence Berkeley National Laboratory, 1 Cyclotron Road, CA 94720, USA}

\author{S. Sinha}
\affiliation{Cornell University, Ithaca, New York, USA}

\author{T. Schenkel}
\affiliation{Acceleration Technology \& Applied Physics, Lawrence Berkeley National Laboratory, 1 Cyclotron Road, CA 94720, USA}

\date{\today}

\begin{abstract}
We report on the development of multi-beam RF linear ion accelerators that are formed from stacks of low cost wafers and describe the status of beam power scale-up using an array of 120 beams. The total argon ion current extracted from the 120-beamlet extraction column was \SI{0.5}{\milli\ampere}. The measured energy gain in each RF gap reached as high as \SI{7.25}{\keV}. We present a path of using this technology to achieve ion currents $>$\SI{1}{\milli\ampere} and ion energies $>$\SI{100}{\keV} for applications in materials processing. 
\end{abstract}

\maketitle

\section{\label{sec:intro} Introduction}
Beams of energetic ions are widely used in material processing and manufacturing \cite{Hamm2011-af} and for testing nuclear materials where the energetic ions mimic damage induced by neutrons.\cite{was2016fundamentals} New directions in particle accelerator development include high gradient normal conducting and superconducting radiofrequency linear accelerators (RF linacs),\cite{Shiltsev2021-oz} laser driven nanostructures \cite{Sapra2020-vo} and plasmas,\cite{Steinke2020-jq} as well as use of techniques from micro-electromechanical systems.\cite{Persaud2017-xz, Seidl2018-xq, Vinayakumar2019-fl, Seidl2018-cw} Today, energetic ions are delivered to targets from accelerators mostly in single beams with ion currents in the range of microampere to a few milliampere. The maximum achievable ion current density is limited by space charge forces during beam transport and the total ion current is limited by the size of the extraction aperture at the ion source where the emittance needs to be kept low enough for efficient transport in the beam line.\cite{Brown2004-eo} The concept of multi-beam ion accelerators was developed in the late 1970s by Maschke \textit{et al.} who coined the term Multiple Electrostatic Quadrupole Linear Accelerator (MEQALAC).\cite{Maschke1979-lx} MEQALACs are radiofrequency(RF)-driven linear accelerators where the total ion current can be scaled by adding more beams and the ion kinetic energy can be increased by adding accelerator modules. In the first implementations, MEQALACs used RF cavities of sizes in the order of centimeters to tens of centimeters to achieve ion acceleration with high voltages driven at frequencies in the \SI{25}{\MHz} range.\cite{Urbanus1989-yu} We have recently reported on the development of multi-beam RF accelerators that we assemble from low cost wafers.\cite{Persaud2017-xz, Seidl2018-xq} RF-acceleration structures and electrostatic quadrupole (ESQ) focusing elements are formed on printed circuit boards (PCB) and silicon wafers with \SI{10}{\cm} diameter using standard microfabrication techniques.\cite{Vinayakumar2019-fl} In our prior work, \SI{0.1}{\milli\ampere} Ar+ ions were extracted through an array of 3×3 beamlets and we have demonstrated beam acceleration at an energy gain of \SI{2.6}{\keV} per acceleration gap.\cite{Persaud2017-xz} In this article, we report on our recent effort in scaling up the beam power by using 120 beamlets with improved energy gain per RF acceleration gap. These advances pave the path towards ion currents $>$\SI{1}{\milli\ampere} and ion energies $>$\SI{100}{\keV} in a compact, low cost setup for applications in material processing and manufacturing.

\section{Beam power scaling}

\subsection{Scaling up the total ion current}
The approach we have taken to scale the ion current to \SI{1}{\milli\ampere} is to use more beamlets. With a \SI{0.5}{\milli\meter}-diameter aperture, Ar+ ion currents on the level of \SI{10}{\micro\ampere} were extracted from each aperture using a filament-driven multi-cusp ion source,\cite{Persaud2017-xz} which indicated an ion current density from the source on the order of \SI{5}{\milli\ampere\per\centi\meter\squared}. With more than one hundred apertures, a total current of \SI{1}{\milli\ampere} is expected using 120 beams with an integrated extraction aperture of \SI{0.24}{\centi\meter\squared}, representing an increase by a factor of over ten corresponding to the larger number of beamlets.\cite{Persaud2017-xz} In the future, ion sources with much higher current densities can also be adapted to beamlet arrays, further boosting the total ion currents.\cite{Ji2016-qn} 
\begin{figure}[htbp]
	\includegraphics[width=0.6\linewidth]{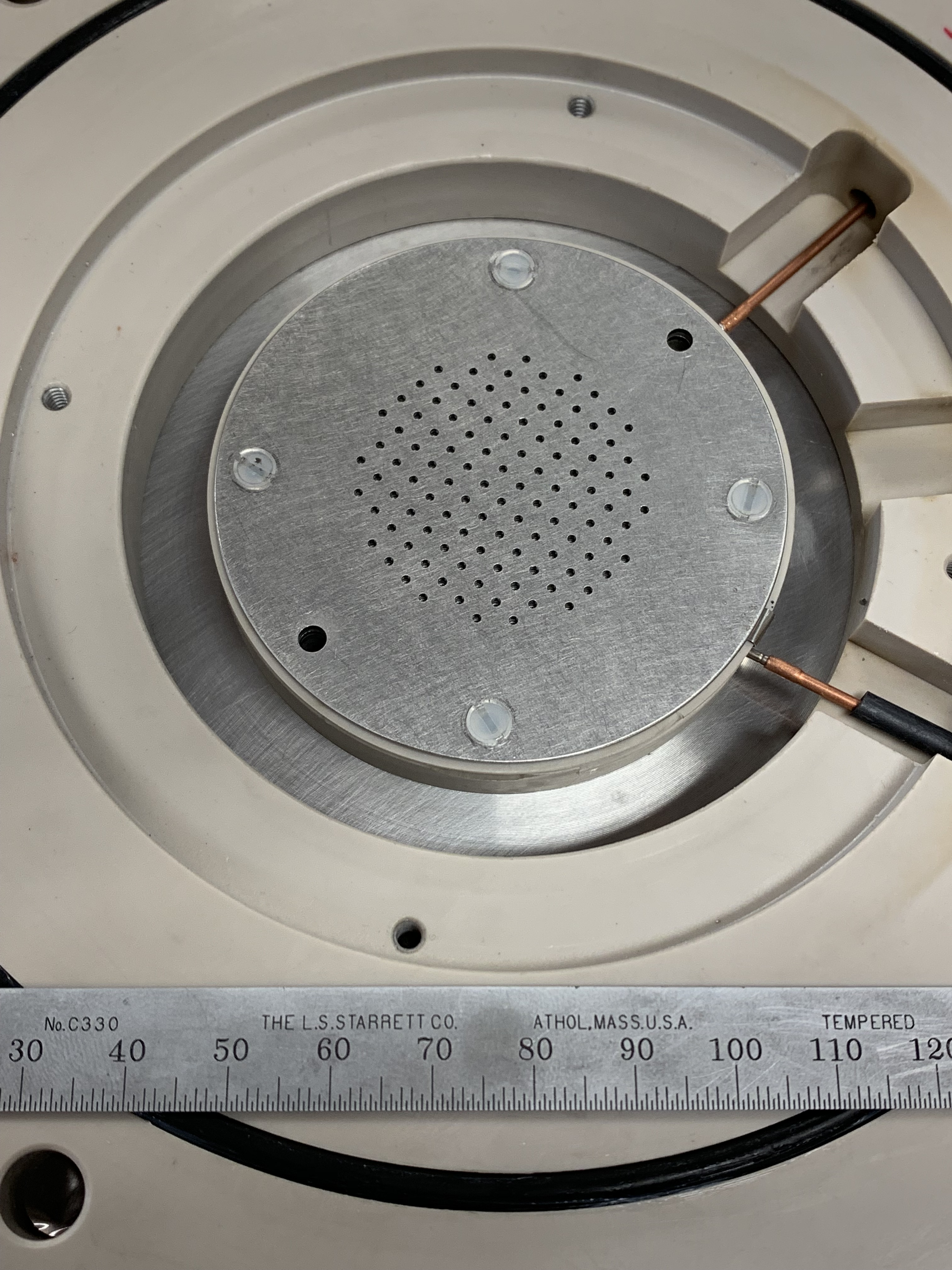}
	\caption{Extraction column with total of 120 beamlets (each of which is \SI{0.5}{\mm} in diameter, and \SI{3}{\mm} center-to-center spacing) in a square array pattern with a diameter of \SI{35}{\mm}.}
	\label{fig:extraction}
\end{figure}
To achieve this goal, a new ion extraction column that consists of a total of 120 beamlets was fabricated, as shown in Fig.~\ref{fig:extraction}. The size of each extraction aperture is \SI{0.5}{\mm} in diameter, and the center-to-center distance is \SI{3}{\mm}. The region that covers all the 120 apertures is about \SI{3.5}{\cm} in diameter. This design is denser than our original 3×3 arrays, where the center to center distance was \SI{5}{\mm}.\cite{Persaud2017-xz}

\subsection{Fabrication of new RF and ESQ wafers with an array of 120 beamlets}
\begin{figure}[htbp]
	\includegraphics[width=\linewidth]{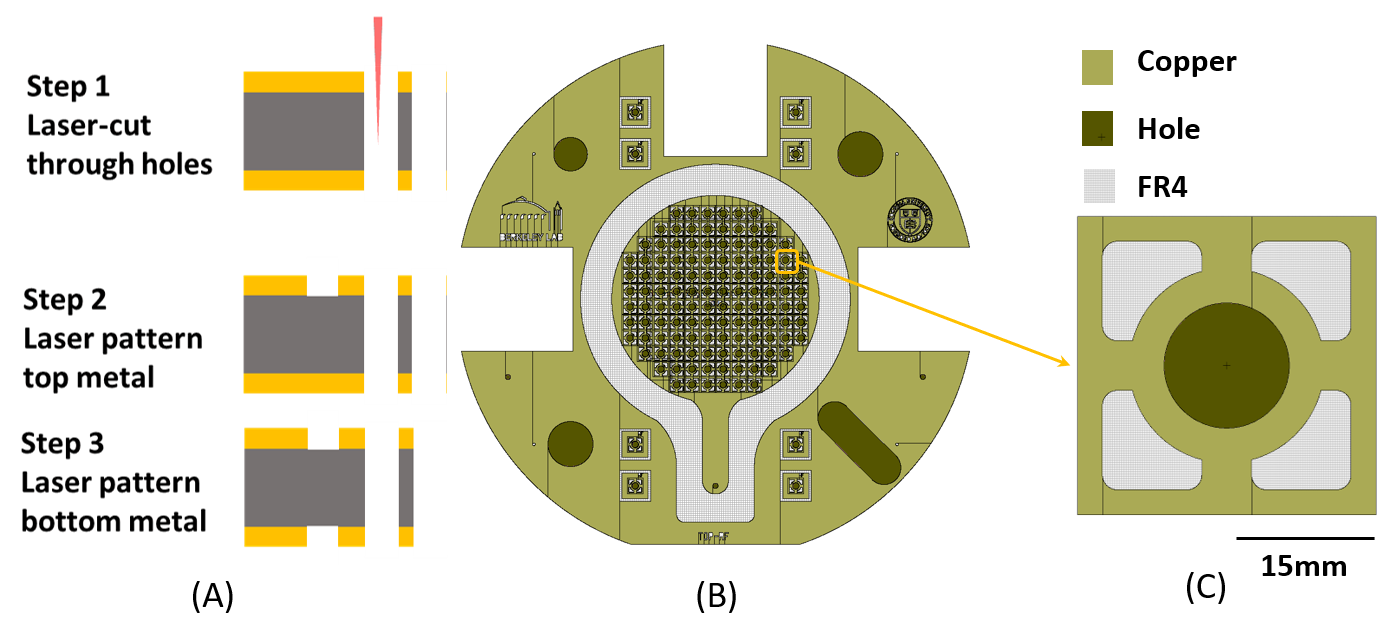}
	\caption{(A) PCB fabrication procedure of RF wafers (B) Layout of RF wafers. The red holes are the laser cut through part and the dark green is places where metal is removed. (C) Design of a unit cell.}
	\label{fig:pcb-procedure}
\end{figure}

\begin{figure}[htbp]
	\includegraphics[width=\linewidth]{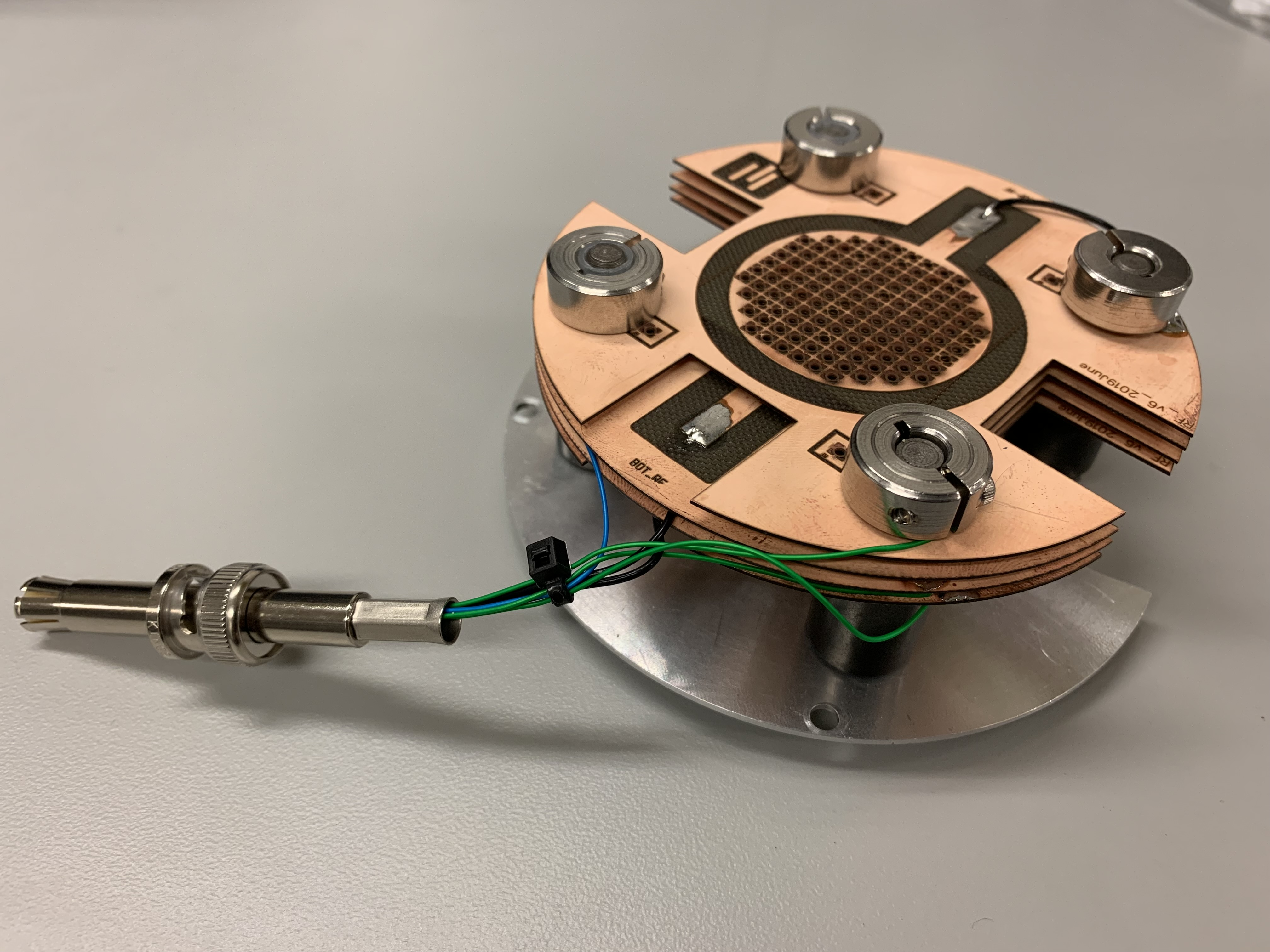}
	\caption{Example of an RF stack with four PCB wafers and 120 beamlet apertures that were manufactured by laser-cutting and assembled for beam experiments.}
	\label{fig:stack}
\end{figure}

In order to accelerate all the ion beamlets extracted from the ion source, RF acceleration and ESQ beam focusing wafers with the same 120-beamlet pattern have been fabricated. 
For RF wafers, we used laser micromachining (LPKF ProtoLaser U4) to pattern the top and bottom metal layers, and to drill holes through the PCB. Alignment between the top and bottom is achieved by using an integrated vision system and prefabricated alignment fiducials. Steps of the process to fabricate RF wafers are given in Fig.~\ref{fig:pcb-procedure}. In this process, we start with a FR-4 based board that has copper cladded on both sides as seen in the cross section (Fig.~\ref{fig:pcb-procedure}(A)). The circular holes are created using a laser tool. Then laser cutting is used to define the top and bottom metal routing. The layout of the fabricated RF wafer is shown in Fig.~\ref{fig:pcb-procedure}(B). Fig.~\ref{fig:pcb-procedure}(C) shows one unit cell of the design. The copper ring has a width of \SI{300}{\micro\meter} and the width of the copper interconnect lines is \SI{500}{\micro\meter}. The decrease of metal area also reduces the capacitive load for each RF acceleration gap, which enables one RF power amplifier to drive multiple RF gaps. A photo of the RF stack with two RF gaps is shown in Fig.~\ref{fig:stack}. The measured capacitive load of this RF unit with each acceleration gap of \SI{2}{\mm} is approximately \SI{23}{\pico\farad}. 

\begin{figure}[htbp]
	\includegraphics[width=\linewidth]{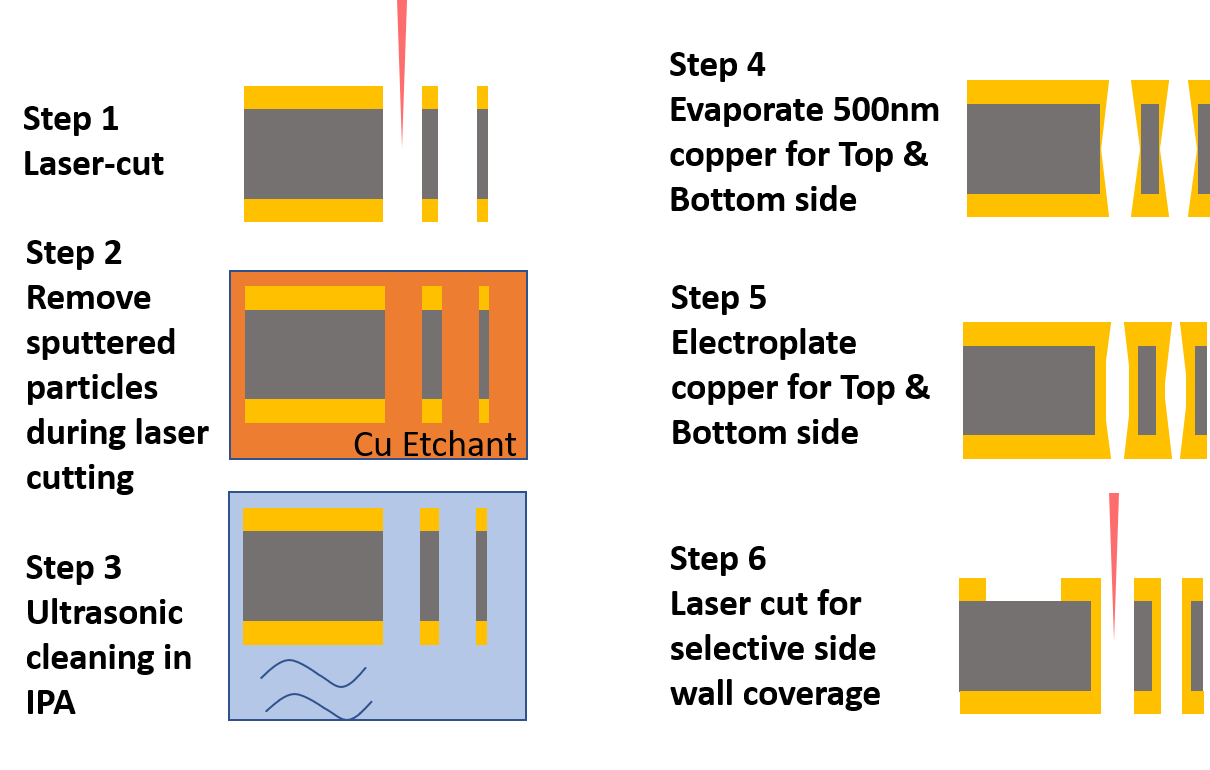}
	\caption{Process flow for ESQ wafer fabrication.}
	\label{fig:process-flow}
\end{figure}

The process flow for ESQ wafer fabrication is depicted in Figure~\ref{fig:process-flow}. The main challenge is to obtain good copper coverage over the selected side walls for each quadrant of the electrodes. The resistance change between top and bottom electrodes has been closely monitored after each processing step. We also included 8 test units outside the focusing array in order to check the sidewall deposition. After evaporation (step 4), the total resistance between top and bottom sides of the wafer was \SIrange{0.1}{0.3}{\ohm}. To obtain more robust sidewall coverage, we added an electroplating process for a thicker copper layer. After electroplating, the total resistance between the top and bottom sides of the wafer was reduced to \SIrange{0.05}{0.1}{\ohm}. We then use an LPKF laser for fine patterning (step 6). 

\begin{figure}[htbp]
	\includegraphics[width=\linewidth]{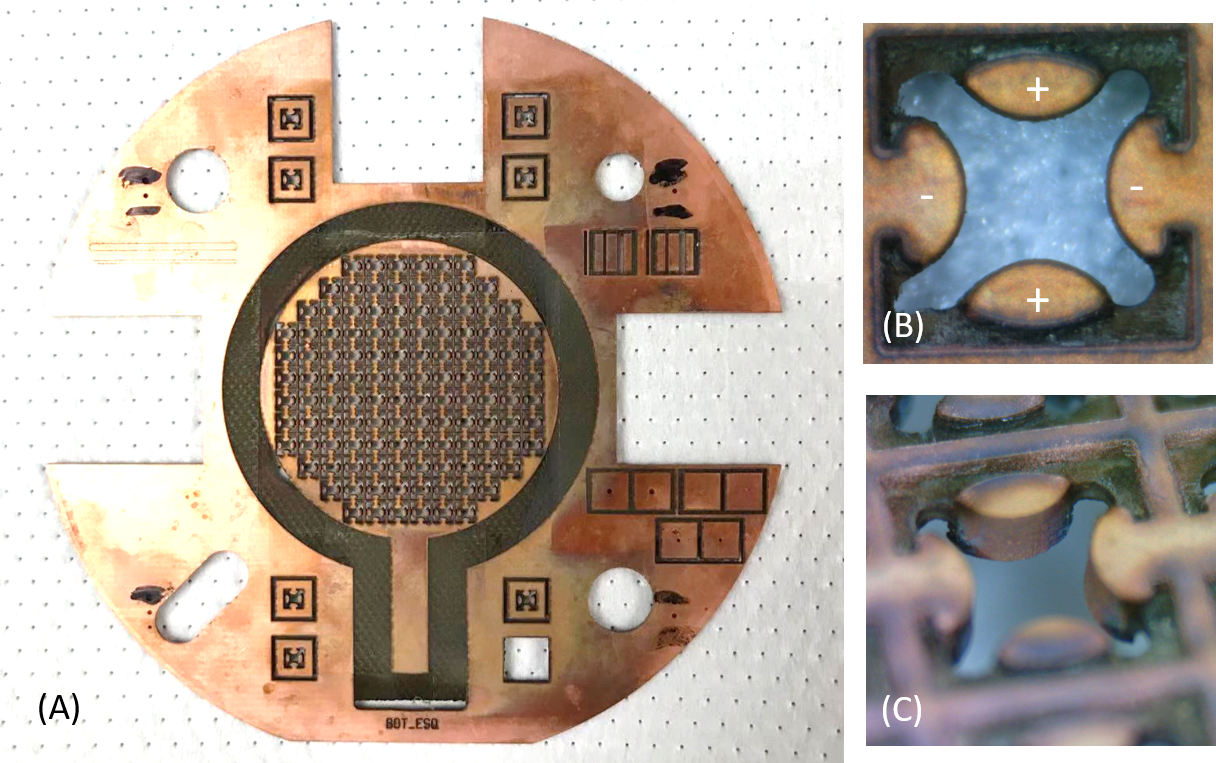}
	\caption{ESQ wafer: (A) ESQ wafer made with a \SI{700}{\micro\meter} thick PCB. (B) a unit cell of the focus unit. (C) Example of a side wall in a unit cell that was well covered with copper.}
	\label{fig:esq}
\end{figure}

The final device with 120 beamlets focus units is shown in Figure~\ref{fig:esq}. A picture of one unit cell is shown on the right in Figure~\ref{fig:esq}, where opposite electrodes are designed to be connected and adjacent electrodes are designed to be isolated from each other. In this device, all facing electrodes have a resistance $<$\SI{1}{\ohm} and all adjacent electrodes have resistances $>$\SI{ 10}{\mega\ohm} between each other. This grants good electrical connections between electrodes with the same electrical potential, as well as good isolation between electrodes with opposite potentials. We tested this wafer up to biases of \SI{500}{\volt} as required for focusing of beams with energies up to a few tens of \si{\keV} and observed low leakage currents of ~\SI{45}{\micro\ampere}.  

\subsection{Scaling up the energy gain per RF acceleration gap}
\begin{figure}[htbp]
	\includegraphics[width=\linewidth]{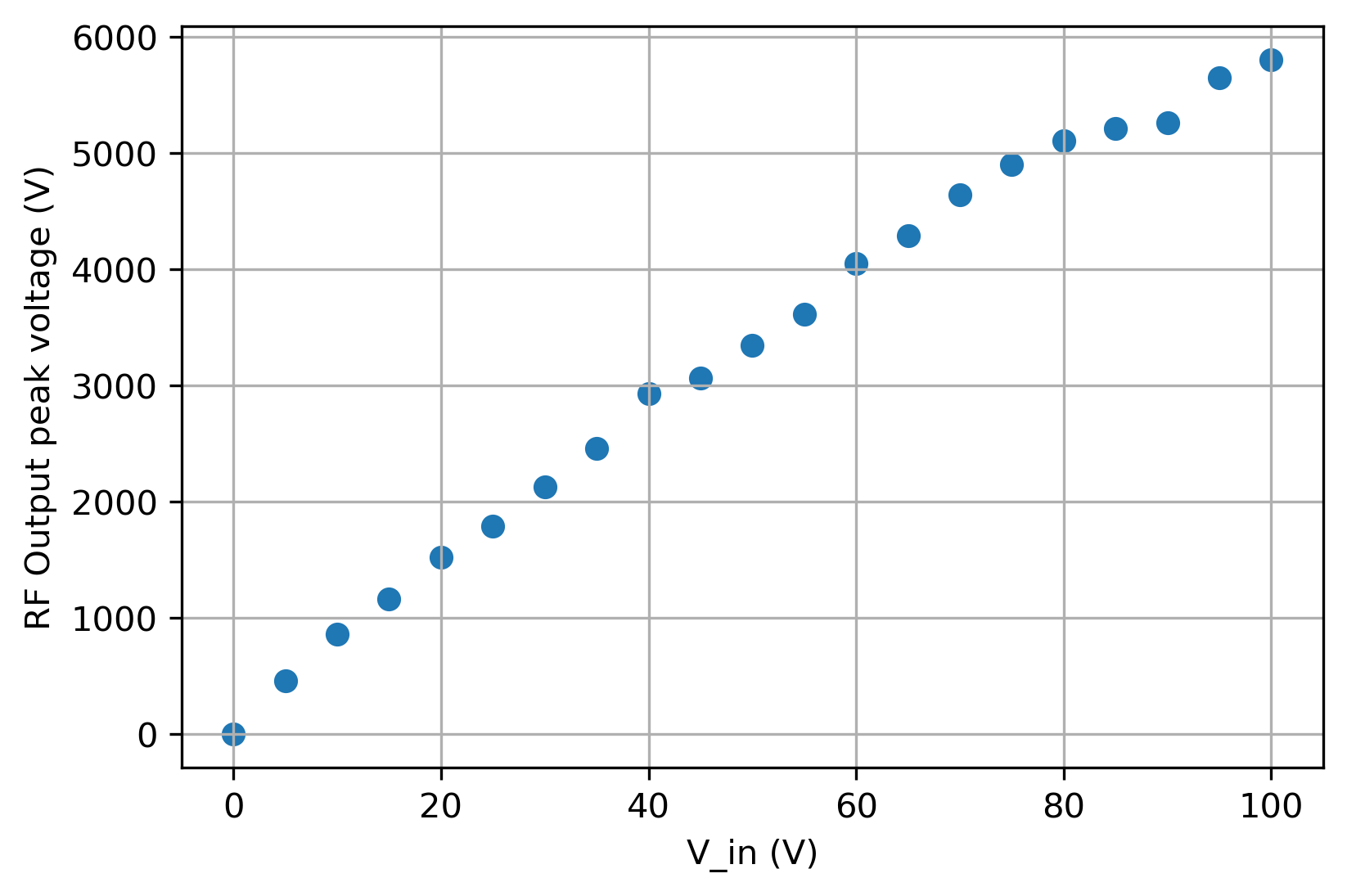}
	\caption{The output peak voltage of the RF amplifier at \SI{14.86}{\MHz}, as a function of the input voltage $V_{in}$.}
	\label{fig:output-voltage}
\end{figure}

The RF high-voltage is generated by an RF power amplifier. Power amplifiers at \SI{13.56}{\MHz} were designed and built by Airity Technology.\cite{Airity}  The resonance frequencies can be slightly adjusted by varying the RF gap distances and the corresponding change in the capacitive loads. The output peak voltage increases linearly as a function of the input voltage, as shown in Figure~\ref{fig:output-voltage}. Based on the output monitor, the peak output voltage was \SI{5800}{\volt} at an input voltage of \SI{100}{\volt}.

\section{Experimental setup}
\begin{figure}[htbp]
	\includegraphics[width=\linewidth]{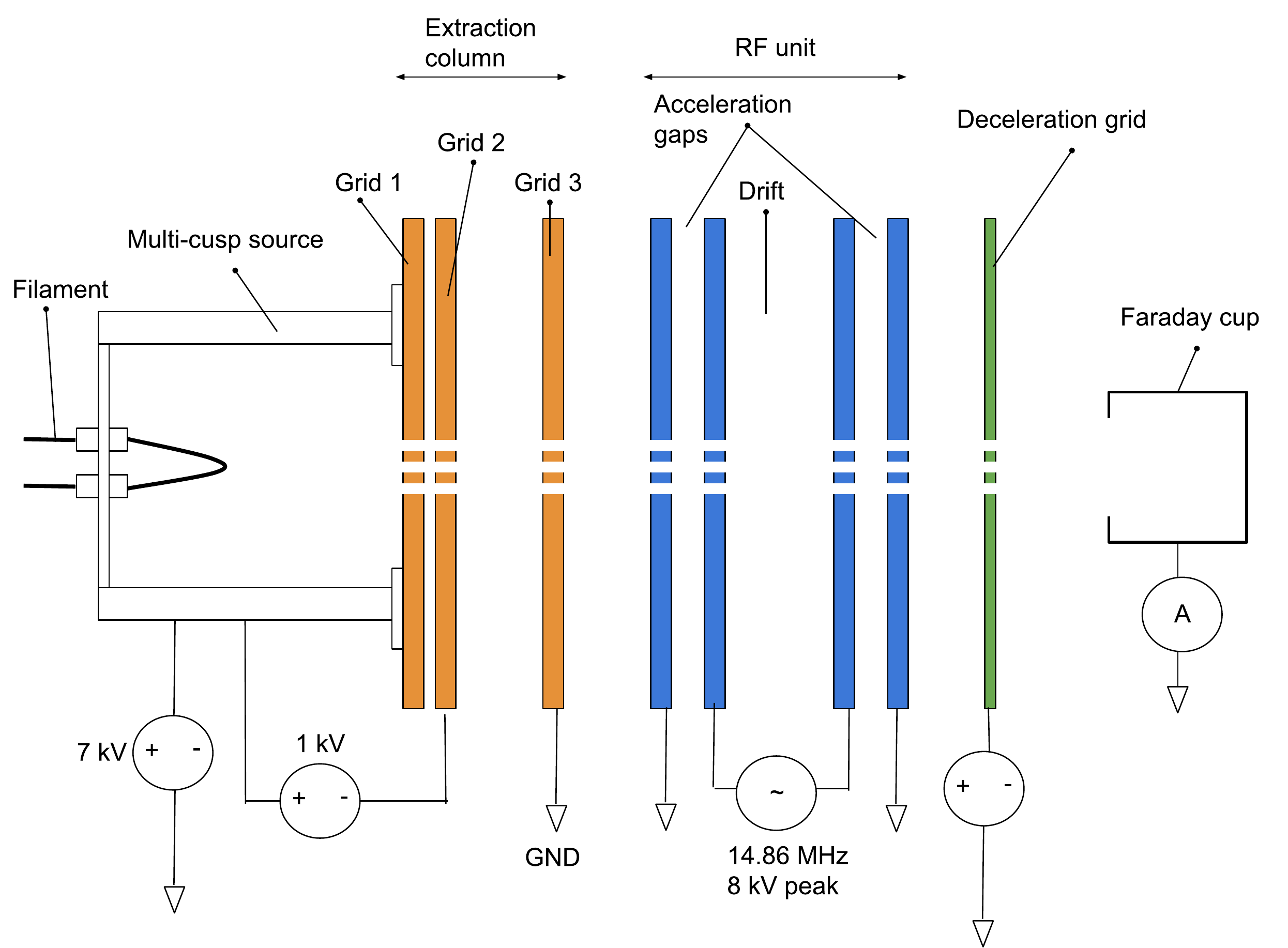}
	\caption{The experimental setup consists of a filament driven ion source with a three grid extraction system that provides an array of 120 beamlets. The source is biased at high voltage for beam extraction. A RF unit cell consisting of two RF gaps is inserted, followed by a Faraday cup to measure the beam current. An energy analyzer can be inserted to measure the ion beam energy profile. The green electrode is the retarding field electrode used as an ion energy diagnostic.}
	\label{fig:setup}
\end{figure}

The experimental setup is the same as described in Ref.~\onlinecite{Persaud2017-xz}.  For our MEMS based multi-beam linac development we have been using a filament driven, multi-cusp ion source.  The source operates using argon (or other gases) in the plasma chamber with an operation pressure of $\sim$\SI{3}{\milli\torr}. A \SI{100}{\volt} arc pulse is applied for \SI{50}{\micro\second} to ignite the plasma. Positive ions are extracted during the arc pulse by floating the source body to a high voltage (\SIrange{7}{10}{\kilo\volt}). The plasma facing electrode (grid 1 in Fig.~\ref{fig:setup}) is not electrically connected to a fixed voltage source and, therefore, floats to the plasma potential during operation. The second electrode (grid 2) is used to extract the ion beam and is biased at a negative voltage relative to the source body. The ions then gain their remaining kinetic energy when they are further accelerated to a grounded exit electrode Grid 3. 

After leaving the beam extraction column, ions are injected into the RF acceleration unit. In this experiment, the test structure consisted of four RF wafers that formed two RF acceleration gaps. The drift distance of 8.4 mm was calculated to match an argon ion beam with an initial beam energy of \SI{7}{\keV}, a driving RF frequency of \SI{14.86}{\MHz}, and an acceleration voltage of \SI{6}{\keV} per gap. 

We operated the RF power amplifier in burst mode. There were $\sim$500 RF cycles in each burst of $\sim$\SI{35}{\us} duration. The ion source delivers a constant ion current over many RF periods. Therefore, ions are accelerated when they arrive in the gap while the RF longitudinal field has a positive polarity.  Ions that do not meet that condition are decelerated or are not accelerated. A retarding potential analyzer is used to measure the beam energy distribution. This is implemented by adding a biased grid after the RF units followed by a Faraday cup to measure the beam current. By scanning the grid voltage, the Faraday cup at the end of the beamline selectively detects the current of ions with a kinetic energy higher than the bias voltage on the grid.

\begin{figure}[htbp]
	\includegraphics[width=\linewidth]{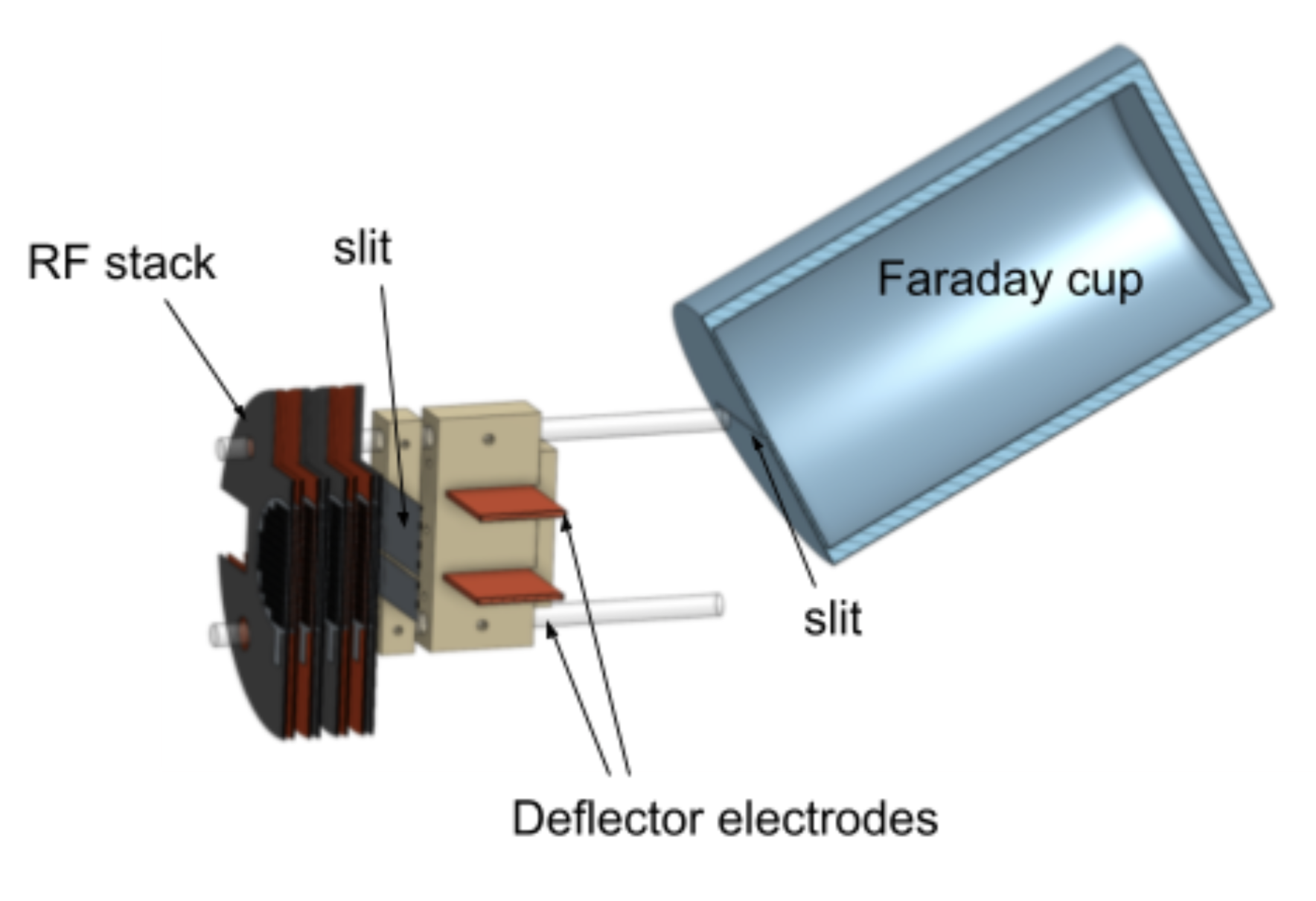}
	\caption{For some beam measurements, two parallel plates biased with HV are installed downstream of the RF stack, instead of the retarding field electrode and Faraday cup diagnostic shown in Fig.~\ref{fig:setup}. The electric field between the plates deflects the ion beams upwards, separating the single ions based on their energy and into a Faraday cup with an entry slit mounted at an angle. In this way the beam current of ions of a specific energy is determined. By varying the HV biases applied on the plates, we can measure the complete energy spectrum to much higher kinetic energy. }
	\label{fig:deflector}
\end{figure}

As we reach higher ion energy using more RF gaps, the retarding potential energy analyzer will no longer be practical due to high voltage breakdown problems, the availability of the high voltage power supply, and potential hazards due to Bremsstrahlung x-ray generation from high energy secondary electrons. We thus transitioned to a new energy analyzer shown in Figure~\ref{fig:deflector} based on deflection of the ions in a parallel plate geometry. A slit installed behind the last RF wafer is used to select only one row of the beamlets. Two parallel plates biased at high voltage are then used to deflect the ion beams. A Faraday cup is mounted at an angle behind the parallel plates to detect ions at a specific deflection angle. By varying the high voltage biases applied on the deflection plates, we can measure the complete energy spectrum of several beamlets.

\section{Results and Discussions}

\subsection{Total beam current extracted from the ion source}
\begin{figure}[htbp]
	\includegraphics[width=\linewidth, clip=true]{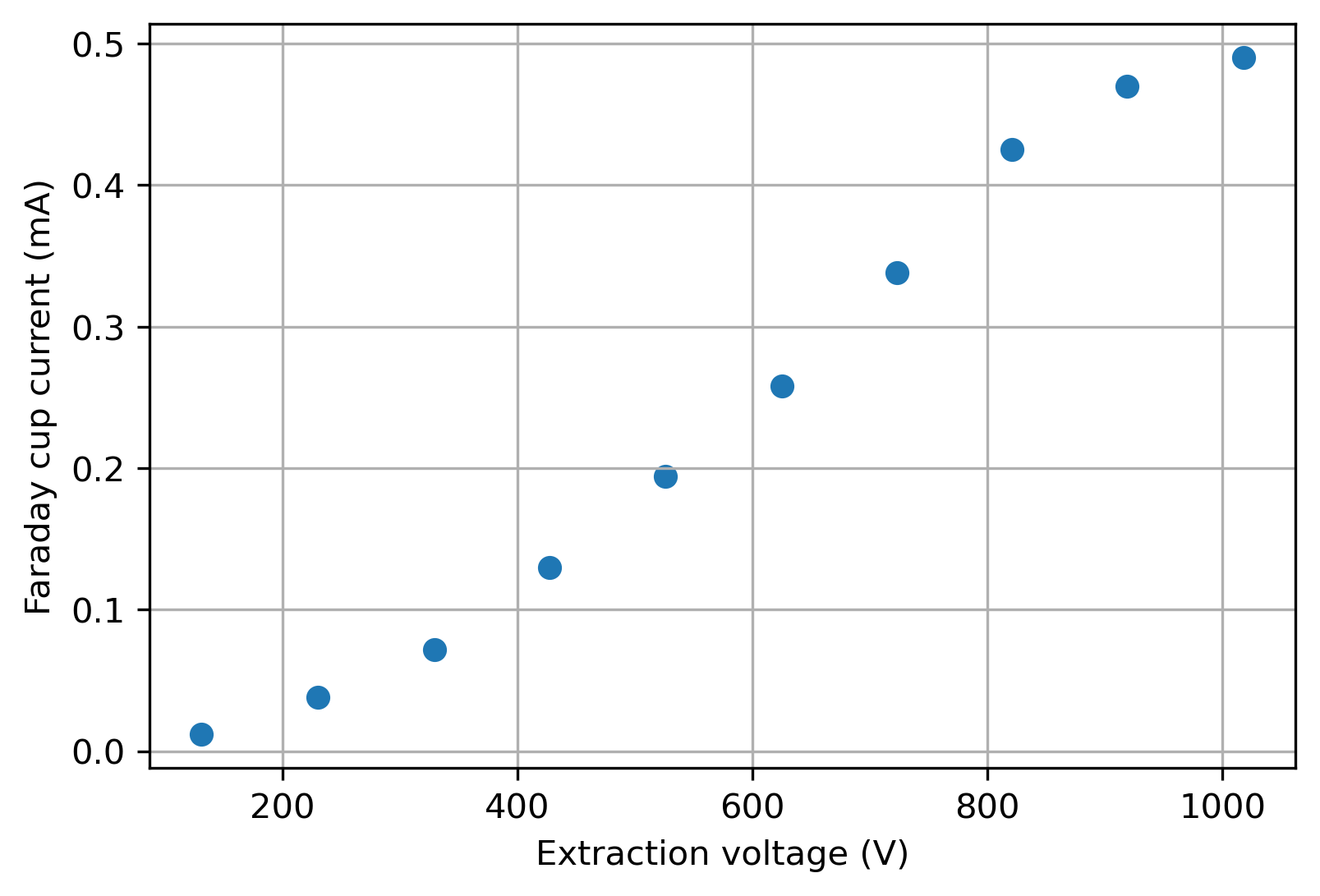}
	\caption{The measured total Ar+ current in the Faraday cup reached \SI{0.5}{\mA} from an array of 120 extraction apertures as shown in Figure~\ref{fig:extraction}.}
	\label{fig:current}
\end{figure}

With the extraction column described above, a current of up to \SI{0.5}{\mA} of Ar+ ions was measured in the Faraday cup, as shown in Fig.~\ref{fig:current}. The total ion current achieved was a factor of two lower than anticipated. One cause may be that the region of uniform plasma in our ion source is smaller than the area covered by the array of 120 beamlets, which has a diameter of \SI{3.5}{\cm} (Figure~\ref{fig:extraction}). 

\subsection{Ion energy distribution measured using a retarding field}
\begin{figure}[htbp]
	\includegraphics[width=\linewidth]{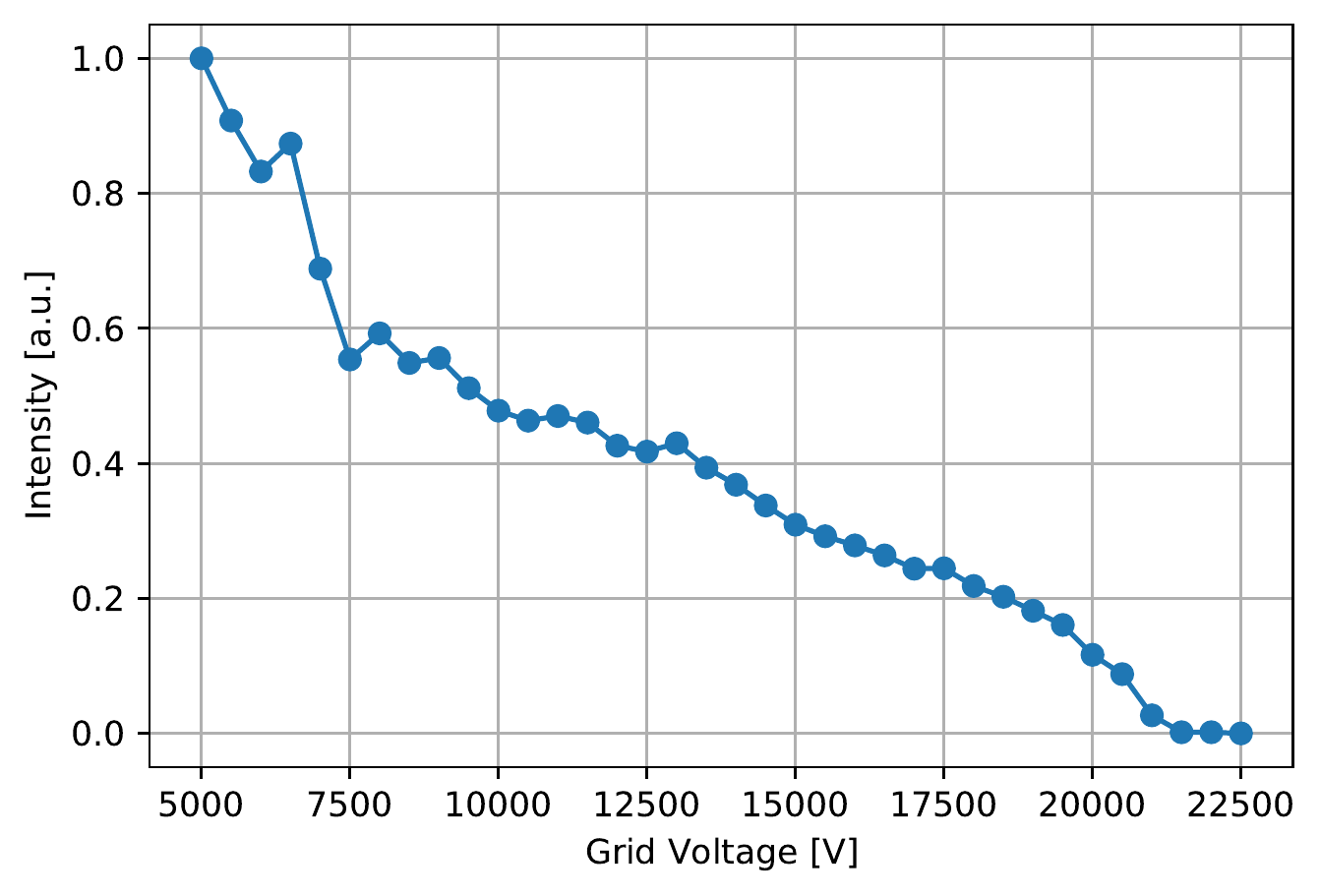}
	\caption{The current of the ions reaching the Faraday cup decreases as the grid biasing voltage increases. With an injection energy of \SI{7}{\keV}, the cut-off energy measured by the energy analyzer is \SI{21}{\keV}, which indicates that the average energy gain per gap is $\sim$\SI{7}{\keV}.}
	\label{fig:grid}
\end{figure}

The beam energy profile was measured with the RF input amplitude set to \SI{100}{\volt}, which resulted in a peak output at approximately \SI{5800}{\volt} based on the capacitor-divider monitoring signal. Using the retarding potential energy filter, the Faraday cup current decreases gradually as the grid bias voltage increases, as shown in Figure~\ref{fig:grid}. It reaches zero when the grid voltage reaches the maximum energy of the incoming ions. With an injection energy of \SI{7}{\keV}, the cut-off ion energy measured by the energy analyzer is \SI{21}{\keV}, which indicates that the average energy gain per gap is $\sim$\SI{7}{\keV}. 

\subsection{Ion energy distribution measured using the parallel-plate energy analyzer}
\begin{figure}[htbp]
	\includegraphics[width=\linewidth, clip=true]{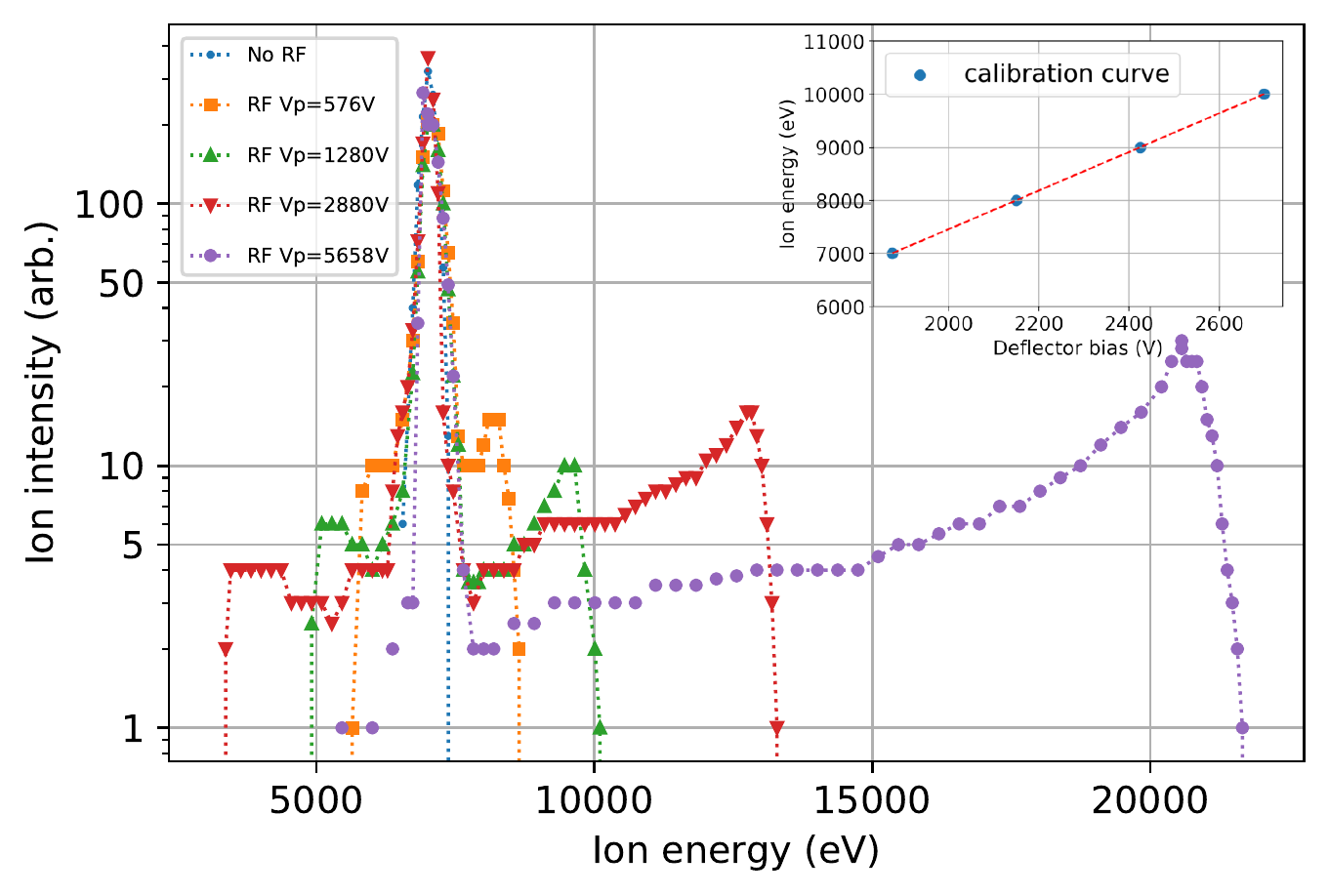}
	\caption{The ion energy distribution after exiting the RF stack at various RF power settings. The maximum energy we reached is approximately \SI{21.5}{\keV} with energy gain of \SI{7.25}{\keV}/gap. The energy analyzer was calibrated using ion beams with different injection energy when no RF acceleration was applied. The calibration curve is shown as the inset plot.}
	\label{fig:energy}
\end{figure}

The parallel-plate energy analyzer was employed to measure the ion energy distribution after being accelerated by the RF stack in Figure~\ref{fig:process-flow}. It was calibrated using monoenergetic ion beams extracted from the ion source at various energies. The inset plot in Figure~\ref{fig:energy} shows the calibration curve. For higher beam energies, a higher bias voltage needs to be applied on the parallel plates to bend the beam so that it can pass through the entry slit at the Faraday cup. The energy of the beam scales linearly with the bias on the parallel plates. With this calibration, we can extrapolate to determine the ion energy when high bias voltages are applied. 
Figure~\ref{fig:energy} depicts the measured ion energy distribution at various RF power settings. At the resonance frequency during operation around 14 MHz, the corresponding RF micro-pulse length is 71 ns, much shorter than the ion beam pulse width of tens of microseconds. Only ions that meet the RF resonant condition are accelerated, others are decelerated or not accelerated at all and transmit the RF stack at the injection energy (\SI{7}{\keV}). With an injection energy of \SI{7}{\keV}, the cut-off energy measured by the energy analyzer when the RF amplifier was running at full power ($V_{in}=\SI{100}{\volt}$) was \SI{21.5}{\keV}, corresponding to an average energy gain in each of the two acceleration gaps of $\sim$\SI{7.25}{\keV}. The results agree with the previous measurement using the retarding field very well. 

\section{Outlook}

We currently operate the MEMS based multi-beam RF linac structures without a buncher, and so only accelerate the ~10\% of ions that arrive at the right phase of the RF cycle.  In future work, we will bunch compress injected ion pulses so that we capture more of the injected beam, which will lead to higher average currents by factors of at least 2 to 5.  In our prior work, we have demonstrated extraction of ions from a similar filament-driven multi-cusp ion source using apertures across a pattern with \SI{6}{\cm} diameter.\cite{Ji2016-qn} A total current up to \SI{150}{\mA} for helium ions was achieved. Adapting an ion source with high current density to this MEMS-based RF accelerator will result in high average beam current as well.

To reach beam energies of \SI{100}{\keV} and beyond the number of acceleration gaps needs to be increased. As more and more RF units are added the length of the accelerator will increase. However, for a fixed RF frequency the effective acceleration gradient will degrade with energies since in our RF design the acceleration gaps need to be spaced with distances, $dx$, given by the resonant condition of $dx = \frac{\beta \lambda}{2}  = \frac{v}{2 f}$, where $v$ is the ion velocity and $\beta=\frac{v}{c}$, $\lambda$ is the RF wavelength, $f$ is the RF frequency, and $c$ is the speed of light. In order to keep the accelerator compact and the acceleration gradient high, these drift lengths need to be kept short. This can be achieved by doubling the RF frequencies as the beam passes through the accelerator. To achieve the goal of a compact accelerator we thus have designed power amplifiers at 27.12 and \SI{54}{\MHz}.  

\begin{figure}[htbp]
	\includegraphics[width=\linewidth]{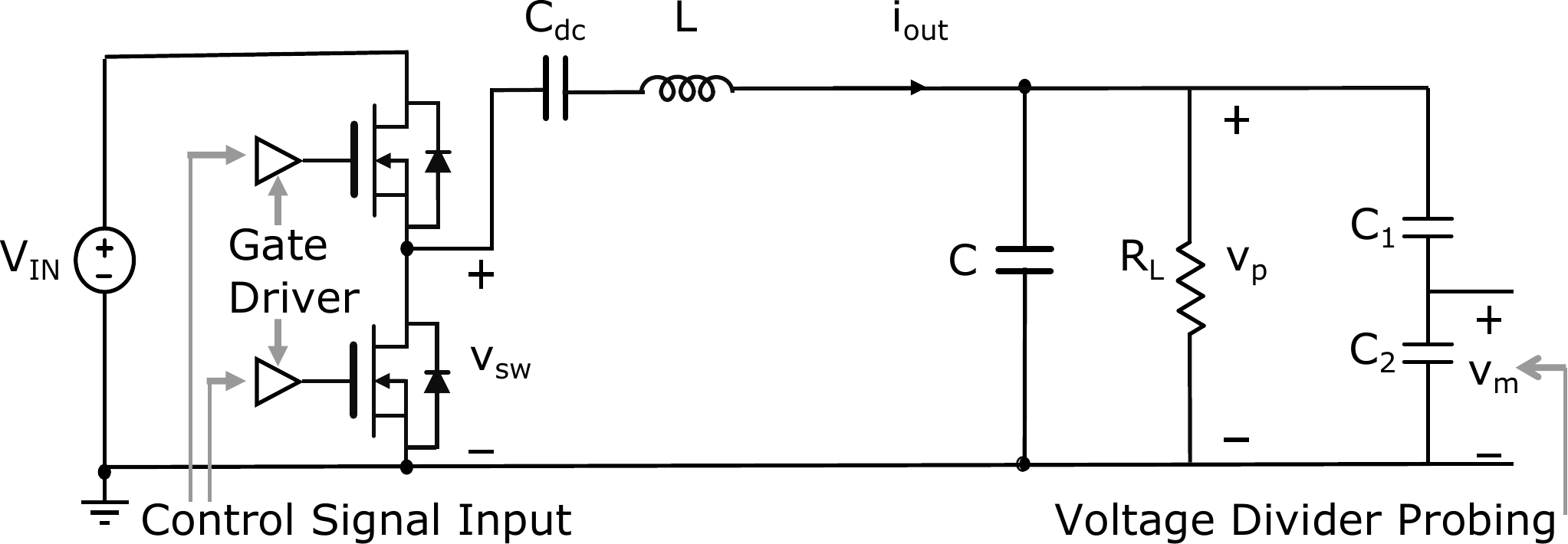}
	\caption{Circuit diagram of a half-bridge class-D power amplifier.}
	\label{fig:class-d}
\end{figure}
Figure~\ref{fig:class-d} illustrates a power amplifier at \SI{27.12}{\MHz} based on a class-D topology. The power amplifier steps voltage by using an LC resonant tank \cite{Network2020-ym}. The power amplifier has the capability to sense the output voltage with low voltage probes, which is enabled by a capacitor voltage divider that scales the output voltage down. The output voltage is measured by using a capacitor divider with a scale-down factor of 470. The measured voltage on the divider has a peak-to-peak value of \SI{14.33}{\volt} which corresponds to \SI{6.9}{\kilo\volt} peak-to-peak voltage on one single RF stack. The power amplifier has a \SI{100}{\volt} dc input voltage. 

\begin{figure}[htbp]
	\includegraphics[width=\linewidth]{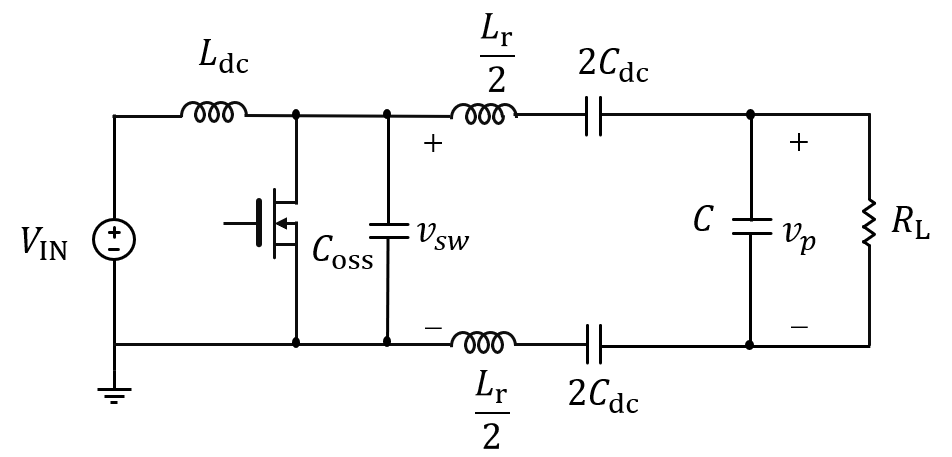}
	\caption{Schematic of the \SI{54}{\MHz} class-E power amplifier.}
	\label{fig:class-e}
\end{figure}
A schematic of the \SI{54}{\MHz} class-E power amplifier\cite{Hou2020-bd} is shown in Figure~\ref{fig:class-e}. The $L_r$ - $C$ resonance is around \SI{54}{\MHz}, and is utilized to achieve high voltage across the plates. A large capacitor $C_{dc}$ is utilized to block the dc component of $v_{sw}$. This single-switch class-E inverter is implemented using a \SI{650}{\volt}/\SI{22}{\ampere} GaN transistor. The coupled inductor is implemented as a single-layer air-core toroid, and a series of high-voltage ceramic capacitors (\SI{10}{\pico\farad}) is utilized to emulate the capacitance of the RF stack. The load presented by the ion beam is emulated using a \SI{2}{\mega\ohm} resistor. Peak output voltage measurements and beam test results will be reported in the future. 

\section{Summary}

We have shown that a new ion extraction column and first RF-based accelerator structure with 120 beamlets has been fabricated, assembled, and tested. The peak output voltage of the RF power amplifier at both \SI{13.56}{\MHz} and \SI{27.12}{\MHz} can reach voltages as high as \SI{7}{\kV}. The total argon ion current extracted from the 120-beamlet extraction column was \SI{0.5}{\mA}. The measured average energy gain per RF gap was approximately \SI{7.25}{\keV}. A new setup to measure the energy distribution of our ion beams using a parallel-plate deflector has been built and shows good agreement with our former method using a retarding grid energy analyzer. The new setup will allow us to measure beams with ion energies up to \SI{100}{\keV}. RF power amplifiers at higher resonance frequencies that can reach output voltages up to \SI{10}{\keV} have been designed. Doubling the acceleration frequency will allow us to keep the overall length of the accelerator short. These are important steps to scale up the beam power of this technology. Our immediate goal is to demonstrate \SI{1}{\mA} of total beam current at ion energies beyond \SI{100}{\keV}. The technology itself can scale to higher beam current and kinetic energy on the order of \SI{1}{\MeV}.

\section*{Author's contributions}
Experiments at LBNL were performed by Qing Ji, Timo Bauer, Grant Giesbrecht, Zihao Qin, and Arun Persaud. Qing Ji, Di Ni, Yuetao Hou, Zihao Qin, Arun Persaud, Peter Seidl, and Thomas Schenkel wrote the manuscript and others edited it. Timo Bauer designed and tested the new energy analyzer. Grant Giesbrecht, Timo Bauer, and Arun Persaud worked on the computer controls.
At Cornell the PCB RF and ESQ wafers were designed and fabricated by Di Ni. Yuetao Hou and Sreyam Sinha designed the 27 and 54 MHz RF amplifiers.

Qing Ji is the worklead of the project. Thomas Schenkel is the principal investigator (PI) of the project and Amit Lal and Khurram Khan Afridi are Co-PIs at Cornell University.

\section*{Data Availability}
The data that support the findings of this study are available from the corresponding author upon reasonable request.

\begin{acknowledgments}
We thank Takeshi Katayanagi for his dedicated and excellent technical support.  We also thank Drs. Luke Raymond, Wei Liang, and Johan Andreasson from Airity Technology for their help on the RF power amplifiers. We acknowledge summer students Madeline Garske and Lindsey Gordon for their contribution to an earlier version of the experiments. This work was funded by ARPA-e. Work at LBNL was conducted under DOE contract DE-AC0205CH11231. Device fabrication was carried out at the Cornell Nano Fabrication (CNF) facility, a member of the National Coordinate Science Foundation (NNCI) network, supported by the National Science Foundation (Grant No. ECCS-1542081).
\end{acknowledgments}

\bibliography{references}
\end{document}